\documentclass[traditabstract]{aa} 
\usepackage{natbib} 
\usepackage{graphicx}
\usepackage{txfonts}
\usepackage{epsfig}
\def\jA{IGR~J08262$-$3736}
\def\jB{IGR~J17348$-$2045}
\def\jC{SAX\,J1818.6$-$1703}
\def\jD{IGR~J17354$-$3255}
\def\jE{IGR~J16328$-$4726} 
\def\inte{{\em INTEGRAL}}
\def\xmm{{\em XMM-Newton}}

\def\chan{{\em Chandra}}

\def\beppo{{\em BeppoSAX}}

\def\swift{{\em Swift}}

\def\ferg{erg/cm$^{2}$/s}

\begin{document}

   \title{\xmm\ observations of four high mass X-ray binaries and \jB\ }

   \author{E. Bozzo 
          \inst{1}
        \and L. Pavan
          \inst{1}
        \and  C. Ferrigno
           \inst{1}   
         \and M. Falanga 
            \inst{2} 
          \and S. Campana
         \inst{3}            
        \and  S. Paltani
          \inst{1}
          \and L. Stella
         \inst{4}
         \and  R. Walter
           \inst{1}
          }
   \institute{ISDC Data Centre for Astrophysics, University of Geneva, Chemin d'Ecogia 16,
             CH-1290 Versoix, Switzerland; \email{enrico.bozzo@unige.ch}
         \and
        International Space Science Institute (ISSI) Hallerstrasse 6, CH-3012 Bern, Switzerland. 
        \and
        INAF - Osservatorio Astronomico di Brera, via Emilio Bianchi 46, I-23807 Merate (LC), Italy. 
        \and
        INAF - Osservatorio Astronomico di Roma, Via Frascati 33, I-00044 Rome, Italy.
             }
   
   \date{Submitted: 2012 January 27; Accepted 2012 July 13}

  \abstract{We present the results of the \xmm\ observations of five hard X-ray emitters: \jA,\ \jD,\ \jE,\ \jC,\ and \jB.\ 
  The first source is a confirmed supergiant high mass X-ray binary, the following two are candidates supergiant fast X-ray transients, \jC\ is a confirmed 
  supergiant fast X-ray transient and \jB\ is one of the still unidentified objects discovered with \inte.\ 
  The \xmm\ observations permitted the first 
  detailed soft X-ray spectral and timing study of \jA\ and provided further support in favor of the association of \jD\ and \jE\ with the supergiant fast X-ray 
  transients. \jC\ was not detected by \xmm,\ thus supporting the idea 
  that this source reaches its lowest X-ray luminosity ($\simeq$10$^{32}$~erg/s) around apastron.
  For \jB\ we identified for the first time the soft X-ray counterpart and proposed the association with a close-by radio object, suggestive of an extragalactic 
  origin.}


\authorrunning{Bozzo et al.}
\titlerunning{XMM-Newton observations of four HMXBs and IGR~J17348$-$2045}

   \maketitle

\section{Introduction}
\label{sec:intro}

High-mass X-ray binaries (HMXBs) comprise a compact object orbiting a massive OB star. The compact object, 
usually a neutron star (NS), emits a conspicuous amount of X-ray radiations 
(up to luminosities of $\sim$10$^{37}$~erg/s) due to the accretion of matter from the OB companion. 
Depending on the nature of the companion star, HMXBs can be divided into Be X-ray binaries (BeXBs) 
and supergiant X-ray binaries (SGXBs). 

In the former, the NS is in a wide eccentric orbit around a Be star. 
The X-ray luminosity is generally low ($\sim$10$^{32}$-10$^{33}$~erg/s) when the NS is far away from 
periastron and accretes matter from low density regions. Remarkable increases in the X-ray luminosity 
($\Delta_{L_{\rm X}}$$\sim$100-1000) are displayed: (i) during the so-called ``Type-II'' X-ray outbursts, 
which last several orbital cycles and present few (if any) X-ray flux variations associated to the orbital phase; 
(ii) at the periastron ( ``Type~I'' outbursts), where the compact object is closer to the companion and accretion takes place 
through the low-velocity and high-density wind of the Be star \citep{stella86,reig11}.  

In the SGXBs, the compact object moves around an early-type supergiant in a nearly circular orbit, and in some cases it is 
well embedded in its dense highly supersonic wind (the so-called ``highly obscured'' SGXBs). The X-ray luminosity is powered by the accretion 
of the strong stellar wind onto the compact star and displays, on-average, less pronounced changes along the orbit with respect to the BeXBs.  
Hydrodynamic instabilities in the wind of the supergiant star can give rise to significant density gradients in the environment 
around the NS, leading to aperiodic variations of the X-ray luminosity ($\Delta_{L_{\rm X}}$$\sim$10-50) on time-scales ranging from 
few to thousands of seconds \citep{negueruela10}. The accretion of particularly dense ``clumps'' in the wind can also induce 
bright short flares that can last for a few hours \citep[see, e.g.,][and references therein]{kreykenbohm11}. 

A few peculiar SGXBs, sharing with the latter a remarkable 
number of similarities \citep[orbital period, energy spectra, properties of the companion stars; see e.g. discussion in][]{bozzo10}, 
were discovered in the late 90s \citep[see e.g.,][]{yamauchi95, smith98}, and 
collectively termed later ``supergiant fast X-ray transients'' \citep[SFXTs;][]{sguera06,negueruela06}. At  
odds with the ``classical'' SGXBs, the SFXTs spend a large fraction of their time \citep{romano11} in a quiescent state with a typical 
luminosity of 10$^{32}$-10$^{33}$~erg/s, and only sporadically undergo bright outbursts ($\Delta_{L_{\rm X}}$$\sim$10$^4$-10$^5$) lasting a 
few hours and reaching peak luminosities comparable with those of the persistent SGXBs \citep[$L_{\rm X}$$\sim$10$^{37}$~erg/s; see e.g.,][]{walter07}. 
The outbursts of SFXTs are associated to the accretion of particularly dense clumps as in other SGXBs, but the origin of the   
lower persistent luminosity and much more pronounced variability of these sources is still a matter of debate \citep{zand05,walter07,grebenev07,bozzo08,bozzo11}. 

In this paper, we report on the \xmm\ observations of four HMXBs discovered with \inte\ and \beppo.\ Among these sources, 
IGR\,J08262$-$3736 is classified as a classical SGXB, IGR\,J17354$-$3255 and IGR\,J16328$-$4726 are candidate SFXTs, and 
SAX\,J1818.6$-$1703 is a confirmed SFXT. We also report on the \xmm\ observation of the still unclassified \inte\ source IGR\,J17348$-$2045. 
A summary of the previous results for all the above sources is presented in Sect.~\ref{sec:sources}, while our analysis and results of the \xmm\ observations 
are presented in Sect.~\ref{sec:data}. Discussions and conclusions on the properties of the soft X-ray emission from the 
five objects are reported in Sect.~\ref{sec:discussion}.

\section{The selected sources}
\label{sec:sources}

\subsection{ \jA}
\label{sec:jA}

\jA\ was discovered by the hard X-ray imager IBIS/ISGRI on-board the \inte\ satellite and reported for the first time in the fourth ISGRI 
catalog \citep{bird10}. The measured fluxes of the source were  $0.4\pm0.1$~mCrab in the 20--40~keV energy band and $0.7\pm0.2$~mCrab in the 
40--100~keV energy range\footnote{Conversions factors are 
$1 \textrm{mCrab}=7.57\times 10^{-12}$~\ferg\ in the 20--40~keV energy band and $1 \textrm{mCrab}=9.42\times 10^{-12}$~\ferg\ in the 40--100~keV energy band,  
respectively. The fluxes of the source thus correspond to $3.0 (\pm 0.8) \times 10^{-12}$~\ferg\ and $7(\pm2) \times 10^{-12}$~\ferg,\ 
respectively.} \citep{bird10}. 

\cite{masetti10} tentatively associated the source to the OB-V star SS~188 located  at a distance of 6.1~kpc, thus suggesting that 
\jA\ was a new member of the HMXB class discovered with \inte.\ The association was later confirmed by 
\citet{maliziaatel} using Swift/XRT data. The best determined XRT position is at 
R.A.(J2000) = 08$^{\mathrm{h}}$26$^{\mathrm{m}}$13.87$^{\mathrm{s}}$ and Dec.(J2000) = -37\degr37\arcmin11.03\arcsec,\ with 
an associated uncertainty of 4~arcsec. The XRT spectrum could be well fit by using an absorbed power-law model with  
an absorption column density of $N_{\rm H}$=1.5$\times$10$^{22}$~cm$^{-2}$ and a power-law photon index of $\Gamma$=2. 
The estimated flux in the 2--10~keV energy band was $\sim$10$^{-11}$~\ferg,\ corresponding to an X-ray luminosity of 
$\sim$2.4$\times$10$^{33}$~erg/s.

\subsection{\jD}
\label{sec:jD}

\jD\ was discovered with \inte\ in 2006 during a monitoring observation of the Galactic bulge region \citep{kuulkers06}, and 
is reported in the fourth IBIS catalogue \citep{bird09} and in the 54 months Swift/BAT hard X-ray catalogue \citep{cusumano10}. The average X-ray 
fluxes in the 20--40~keV and 15--150~keV energy band are 1.1$\times$10$^{-11}$~erg/cm$^2$/s and 2.1$\times$10$^{-11}$~erg/cm$^2$/s, respectively. 
The long term monitoring of \jD\ carried out with the \inte\,/ISGRI and \swift\,/BAT suggested that the source 
is a weak persistent emitter in the hard X-rays (average flux of 1.1~mCrab in the 18--60~keV energy band), only sporadically 
displaying relatively short flares with duration from few hours to $\sim$1~day. This behaviour, together with the periodic modulation detected 
in the hard X-ray data at 8.4474$\pm$0.0017~days, led to the conclusion that \jD\ is a HMXB, possibly 
a SFXT \citep{dai11,sguera11}. The source is also positionally coincident with the high energy AGILE transient AGL\,J1734$-$3310, even though 
the localization uncertainties are still too large to claim a firm association \citep{bulgarelli09}. 

In the soft X-ray domain, the source was observed twice with \swift\,/XRT and once with \chan.\ In the most recent XRT observation \citep{vercellone09}, two sources were found within 
the \inte\ error circle of \jD.\ The first object (S1) displayed a higher flux (1.1$\times$10$^{-11}$~erg/cm$^2$/s, 0.3--10~keV) and a spectrum characterized 
by a blackbody temperature of $kT$$\sim$1.4~keV and an absorption column density of $N_{\rm H}$$\sim$5$\times$10$^{22}$~cm$^{-2}$. The second source (S2) was a factor of 
$\sim$20 fainter and no spectral information was available. In a previously performed XRT observation (2008 March 11), only S2 was detected 
at a nearly constant flux, while S1 was not detected with a 3~$\sigma$ upper limit on its X-ray flux a factor $\simeq$30 lower than the 
one measured in the other XRT observation \citep{vercellone09}. 

The presence of these two sources was confirmed also through the 
\chan\ observation performed on 2009 February 6 \citep{tomsick09}. S1 (=CXOU~J173527.5$-$325554) was detected with an unabsorbed 0.3--10~keV flux of 1.3$\times$10$^{-11}$~erg/cm$^2$/s, 
and displayed an X-ray spectrum characterized by $N_{\rm H}$$\sim$7.5$\times$10$^{22}$~cm$^{-2}$ and $\Gamma$$\simeq$0.54. 
The \chan\ refined position of this source (R.A.(J2000)=17$^{\mathrm{h}}$35$^{\mathrm{m}}$27.59$^{\mathrm{s}}$ 
Dec.(J2000)=-32\degr55\arcmin54.4\arcsec,\ associated 90\% c.l. uncertainty 0.64\arcsec) led to the identification of the infrared counterpart  
2MASS\,J17352760$-$3255544. No catalogued optical counterpart was found at the \chan\ position, consistent with the high measured extinction in the direction of the source. 
In the \chan\ observation, S2 (=CXOU~J173518.7$-$325428) was detected with an X-ray flux of 1.4$\times$10$^{-12}$~erg/cm$^2$/s and displayed an absorbed power-law shaped 
spectrum with $N_{\rm H}$$\sim$2.6$\times$10$^{22}$~cm$^{-2}$ and $\Gamma$$\simeq$0.79. The best estimated source position was 
$R.A.(J2000)=17^{\mathrm{h}}35^{\mathrm{m}}18.73^{\mathrm{s}}$, 
$Dec.(J2000)=\mbox{-32\degr54\arcmin28.7\arcsec}$,\ with an associated 90\% c.l. uncertainty of 0.64\arcsec.\

The more pronounced variability of S1 led to the conclusion that this source was the most likely counterpart to 
\jD\ \citep{vercellone09,tomsick09}.

\subsection{\jE}
\label{sec:jE}

The source \jE\ was discovered with \inte\ by \citet{bird07} and reported as a hard X-ray transient 
in the latest ISGRI and \swift\,/BAT catalogues \citep{bird10,cusumano10}. 
A relatively bright outburst from the source was detected with the \swift\,/BAT on 2009 June 9 \citep{grupe09}. 
The follow-up observations with the narrow field instrument on-board \swift,\ XRT, showed that the spectrum of the source 
could be well described by an absorbed power-law model with $N_{\rm H}$$\simeq$8$\times$10$^{22}$~cm$^{-2}$ and $\Gamma$$\simeq$0.6. 
The estimated 0.3--10~keV X-ray flux was 2.4$\times$10$^{-10}$~\ferg.\ On that occasion, XRT followed the evolution of the 
X-ray flux up to 4~days after the onset of the outburst \citep{fiocchi10}. During the last observation performed on 2009 June 14 the source was no longer  
detected, resulting in a 3$\sigma$~upper limit on the X-ray flux of 1.4$\times$10$^{-12}$~erg/cm$^2$/s \citep[0.3--10~keV;][]{fiocchi10}. 

From the analysis of the BAT data, \citet{corbet10} revealed a highly significant modulation at $\sim$10~d. This was interpreted as 
the orbital period of a HMXB, possibly a supergiant system.  

An accurate analysis of all the archival \inte\ observations performed in the direction of \jE\ was reported by \citet{fiocchi10}. 
In the hard X-ray domain (20--50~keV), two outbursts lasting a few hours were detected by the hard X-ray imager ISGRI on-board \inte.\ No simultaneous data were 
available with the two X-ray monitors JEM-X. On both occasions the ISGRI spectrum could be well fit by using a simple power-law model with 
$\Gamma$$\simeq$2-2.6 and a flux of 2-3.3$\times$10$^{-10}$~erg/cm$^2$/s (20--50~keV). 
The 3$\sigma$ upper limit on the source emission outside the outbursts derived from the ISGRI data was 2.5$\times$10$^{-12}$~erg/cm$^2$/s  
(20--50~keV). 

The fast flaring behaviour of the source, and its spectral properties, suggested that \jE\ is a member of the SFXT class  
discovered with \inte.\

\subsection{\jC}
\label{sec:jC}

\jC\ is one of the confirmed SFXT sources, and was discovered by \beppo\ on 1998 March 11 during a period of intense 
X-ray activity that lasted for about 2~h and reached a peak flux of $\sim$400~mCrab 
\citep[9--25~keV,][]{zand98}.  Since then, several outbursts from the source 
were recorded with \inte\ and \swift\ \citep[see, e.g.][and references therein]{sidoli09}. 
The best X-ray source position was provided by \chan\ at  
R.A.(J2000)=18$^{\rm h}$18$^{\rm m}$37\fs89 and 
Dec.(J2000)=\mbox{-17}\degr02\arcmin47.9\farcs \citep[associated uncertainty 0.6'' at 90\% 
confidence level;][]{zand06}. The estimated source distance is 2.1$\pm$0.1~kpc 
\citep{torrejon10}. 
An in-depth study of the X-ray activity of the source was carried out by 
\citet{bird09} and \citet{zurita09}. The first investigation determined the best orbital period of \jC\ at 
30$\pm$0.1~d \citep[in the following we also consider the reference phase~0 at 53671~MJD, as reported by][]{bird09}. 
\citet{zurita09} also found that most of the discovered outbursts took place when 
the NS passed close to the periastron, and that the source usually 
remains relatively bright in X-rays (a few times 10$^{-11}$~erg/cm$^{2}$/s) 
for about $\sim$6~d around this orbital phase.  
However, outbursts in several periastron passages were missing. This behaviour could be reasonably well 
explained by assuming that the NS in \jC\ has an eccentric orbit ($e$=0.3-0.4) and is sporadically accreting 
mass from the clumpy wind of its supergiant companion (see also Sect.~\ref{sec:intro}). 
On 2006 October 6 \jC\ was also observed with \xmm\ for the first time close to the apastron passage 
\citep[phase 0.51;][]{bird09}, but was not detected. \citet{bozzo08c} determined a 3$\sigma$ upper limit on its unabsorbed 
X-ray flux of 1.1$\times$10$^{-13}$~erg~/cm$^{2}$/~s (the effective exposure time was $\sim$13~ks).

\subsection{\jB}
\label{sec:jB}

\jB\ is an unclassified source discovered with \inte\ and reported for the first time by \citet{bird07}. 
The best determined position of the source with ISGRI is at 
$R.A.(J2000)=17^{\mathrm{h}}34^{\mathrm{m}}57^{\mathrm{s}}$, 
$Dec.(J2000)=\mbox{-20\degr44.8\arcmin}$, with a 90\% c.l. associated uncertainty of 3.6~arcmin \citep{bird10}.  
The estimated fluxes in the 20--40~keV and 40--100~keV energy bands are $0.3\pm0.1$ mCrab and $0.9\pm0.1$~mCrab, respectively.

\section{Data analysis and results}
\label{sec:data}

For the present study we used all the available \xmm\ observations of the five sources   
obtained through the guest observation programs awarded to our research group.  

\xmm\ observation data files (ODFs) were processed to produce calibrated
event lists using the standard \xmm\ Science Analysis
System (v. 11.0). We used the {\sc epproc} and {\sc
emproc} tasks to produce cleaned event files from the EPIC-pn and MOS
cameras, respectively. EPIC-pn and EPIC-MOS event files were filtered in the 0.5--12~keV and 
0.5--10~keV energy range, respectively, to exclude high background time intervals.  
The effective exposure time for each observation and camera is specified for each source  
in the following sections. Lightcurves and spectra for the five sources and the relative backgrounds 
were extracted by using regions in the same CCD.  The difference in extraction areas
between source and background was accounted for by using the SAS {\sc
backscale} task for the spectra and {\sc lccorr} for the lightcurves. 
All EPIC spectra were rebinned before fitting in order to have at least 25 counts per bin and, at the same 
time, to prevent oversampling of the energy resolution by more than a factor of three. 
Where required, we barycenter-corrected the photon arrival times in the EPIC event files 
with the {\em barycen} tool.
Throughout this paper, uncertainties are given at 90\%~c.l., unless stated otherwise.

\subsection{ \jA\ }
\label{sec:jA-xmm}

\begin{figure}
\centering
\includegraphics[scale=0.36,angle=-90]{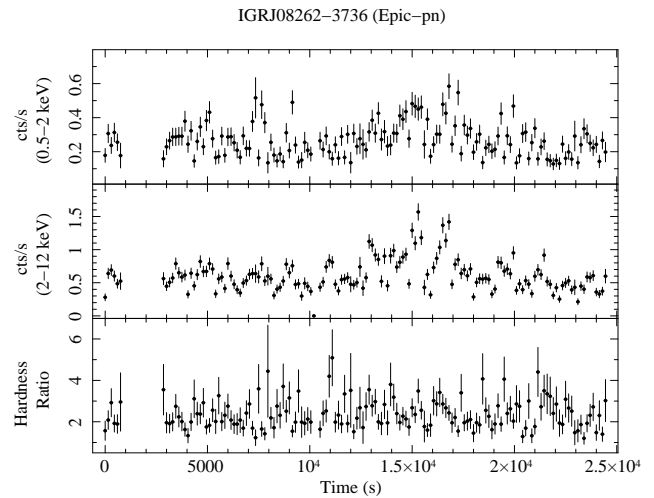}
\caption{ \xmm\ background-subtracted lightcurve of \jA\ in two energy bands and the corresponding hardness ratio. 
The latter is calculated as the ratio between the source count rate in the 2--12~keV and 0.5--2~keV  
energy bands. The time bin is 150~s.}    
\label{fig:jAlcurve} 
\end{figure}

\begin{figure}
\centering
\includegraphics[scale=0.36,angle=-90]{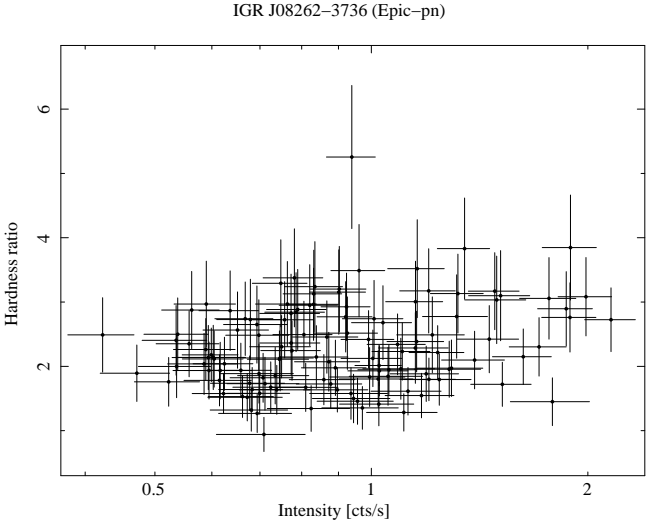}
\caption{Hardness-intensity diagram of \jA\ obtained by using EPIC-pn data. For this plot, 
data were rebinned in order to have S/N$\gtrsim$5 in each bin.}     
\label{fig:jAHID} 
\end{figure}

\xmm\ observed the field-of-view (FOV) around \jA\ on 2010 October 16 for a total exposure time of $\sim$25~ks.
All three EPIC cameras were operated in full frame mode.
A bright point source was detected at a position (obtained from the {\sc edetect$\_$chain} tool)  
R.A.(J2000)=08$^{\rm h}$26$^{\rm m}$13\fs7 and Dec.(J2000)=-37\degr37\arcmin11\farcs58 
\citep[the nominal positional accuracy of the EPIC cameras is assumed 
hereafter to be $\sim$2 arcsec unless otherwise stated; see discussion in][]{pavan11}, 
compatible with that of the soft X-ray counterpart 
of \jA\ identified by \citet{maliziaatel}. On the EPIC-pn, the source lied close to the border between two CCDs 
and thus we chose the extraction region in order to avoid any contribution from the CCD 
rim\footnote{At the time of the \xmm\ observation, the refined XRT position of 
\jA\ \citep{maliziaatel} was not available.}. 
The source lightcurve in two energy bands is shown in Fig.~\ref{fig:jAlcurve}, together with the hardness ratio. 
The source displayed a moderate variability, with two relatively small flares occurring about 1.5$\times$10$^4$~s 
after the beginning of the observation. The hardness ratio and the hardness intensity diagram of the source 
(see Fig.~\ref{fig:jAlcurve} and \ref{fig:jAHID}) did not show clear evidence of variations in the source spectral properties 
with the count-rate. 
\begin{figure}
\centering
\includegraphics[scale=0.4,angle=-90]{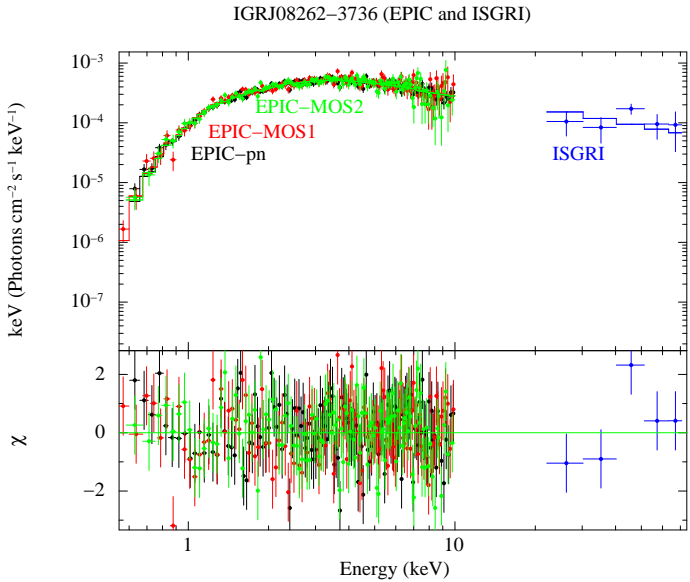}
\caption{Average \xmm\,/EPIC and \inte\,/ISGRI spectra of \jA.\  
The best-fit model (solid line) is obtained with a partial-covering power-law model. The residuals from the fit 
are shown in the bottom panel.}    
\label{fig:jAspectrum} 
\end{figure}

An acceptable fit to the EPIC-pn spectrum of the source accumulated over  
the entire exposure time available could not be obtained using a simple 
absorbed power-law model ($\chi^2_{\rm red}$/d.o.f.=1.7/134). The fit was significantly improved by adding 
a blackbody component at the lower energies (hereafter, BB) or by the introduction of a 
partial covering model. In order to constrain better the spectral 
parameters, we also extracted the spectra from the EPIC-MOS cameras and the average \inte\,/ISGRI spectrum of \jA\ (total effective exposure time 1.2~Ms) 
from the {\sc Heavens} on-line tool\footnote{http://www.isdc.unige.ch/heavens.}.  
The combined EPIC+ISGRI spectra of the source are shown in Fig.~\ref{fig:jAspectrum}. The results of all the fits 
are reported in Table~\ref{tab:jAspec}. A normalization constant with respect to the EPIC-pn was included in all the fits to take into account 
the inter-calibration between the instruments and the possible variability of the source. 
No significant evidence for coherent modulations was found in the EPIC data of \jA.\  
\begin{table*}
\centering
\caption{Results of the spectral fit parameters obtained from the average EPIC+ISGRI spectra of \jA\ (see Sect.~\ref{sec:jA-xmm}).}
\scriptsize
\begin{tabular}{@{}lllllllllllll@{}}
\hline
\hline
\noalign{\smallskip} 
 Model       &  $N_{\rm H}$ & $\Gamma$ & $N_{\rm Hb}$ & $f$ & $kT_{\rm BB}$ & $R_{\rm BB}$ & $F_{\rm 0.5-2~keV}$ & $F_{\rm 2-10~keV}$ & $F_{\rm 20-40~keV}$ &  C$_{\rm MOS1}$ (C$_{\rm MOS2}$)  &  C$_{\rm ISGRI}$ & $\chi^2_{\rm red}$/d.o.f. \\
\noalign{\smallskip} 
\hline        
\noalign{\smallskip} 
1 & 1.4$\pm$0.1 & 1.52$\pm$0.05 & --- & --- & 0.10$\pm$0.01 & 132$_{-33}^{+11}$ & 16.0 & 6.0 & 4.0 &  1.03$\pm$0.03 (1.01$\pm$0.03) &  0.7$\pm$0.2 &1.1/380 \\
\noalign{\smallskip} 
                & 0.60$\pm$0.07 & 1.3$\pm$0.1 & --- & --- & 1.2$\pm$0.2 & 0.18$\pm$0.03 & 1.0 & 5.3 & 3.6 & 1.00$\pm$0.03 (1.00$\pm$0.03) & 0.6$_{-0.2}^{+0.4}$ &0.9/380 \\
\noalign{\smallskip}
2 & 0.80$\pm$0.06 & 1.8$\pm$0.1 & 3.8$\pm$0.7 & 0.63$\pm$0.04 & --- & --- & 4.5 & 6.5 & 4.4 & 1.00$\pm$0.03 (1.00$\pm$0.03) & 1.2$\pm$0.4 &0.9/380 \\
\noalign{\smallskip}
\hline
\end{tabular}
\label{tab:jAspec}
\tablefoot{The leftmost column indicates the spectral model used in {\sc Xspec}. Model 1 is phabs*(BB+pow), model 2 is phabs*pcfabs*pow 
({\sc phabs} is the absorption component, {\sc pow} the powerlaw, {\sc BB} the blackbody, 
and {\sc pcfabs} the partial covering). We indicated with 
$\Gamma$ the power-law photon index, $N_{\rm H}$ the absorption column density in units of 
10$^{22}$~cm$^{-2}$, $f$ the covering fraction in the partial covering model (the local absorption component is indicated with 
$N_{\rm Hb}$ in units of 10$^{22}$~cm$^{-2}$), and $kT_{\rm BB}$ ($R_{\rm BB}$) the temperature (radius) of the BB component 
in keV (km). For $R_{\rm BB}$ we considered a source distance of 6.1~kpc (see Sect.~\ref{sec:intro}). 
For the model that includes the BB component (Model 1), we report both the best fit values 
obtained with a hotter ($kT$$\sim$1~keV) and colder ($kT$$\sim$0.1~keV) emitting surface.  
We also reported the normalization constants C$_{\rm MOS1}$, C$_{\rm MOS2}$, and C$_{\rm ISGRI}$ for 
the MOS1, MOS2 and ISGRI spectra, respectively (we fixed the EPIC-pn constant to 1 as a reference). 
The unabsorbed fluxes (units of 10$^{-12}$~erg/cm$^2$/s) in the 0.5--2~keV, 2--10~keV, and 20--40~keV are indicated 
in all cases with $F_{\rm 0.5-2~keV}$, $F_{\rm 2-10~keV}$, and 
$F_{\rm 20-40~keV}$, respectively.}
\end{table*}

\subsection{ \jD\ }
\label{sec:jD-xmm}

The FOV around \jD\ was observed with \xmm\ on 2011 March 6 for a total exposure time of 20~ks. 
The EPIC-MOS1 and EPIC-pn cameras were operated in full-frame, while the EPIC-MOS2 was in small-window mode.
No significant time intervals were affected by a high flaring background, and thus in the following analysis 
we retained the entire exposure time available for the three instruments. 

Among the two candidate soft X-ray counterparts of \jD\ identified previously with \chan\ (see Sect.~\ref{sec:jD}),  
S1 was not detected by the three EPIC cameras. 
We calculated an upper limit on the emission from the source by using the \verb+ximage+ tool \emph{uplimit}, and 
a circular extraction region of 15~arcsec centered on the \chan\ position of the source (see Sect.~\ref{sec:jD}). 
We obtained a 3~$\sigma$ upper limit on the source count-rate of 0.002 cts/sec in the 0.5--10~keV energy band 
(effective exposure time 19.0~ks). This corresponds to an observed X-ray flux of 7$\times$10$^{-14}$~\ferg 
\citep[we assumed a spectral index of $\Gamma$=1.7 and an absorption column density of $N_{\rm H}$=7$\times$10$^{22}$~cm$^{-2}$ 
as reported by][]{dai11}. 

S2 was outside the EPIC-MOS2 FOV, but clearly detected by the EPIC-pn and the MOS1. In the former instrument the source was 
located on the gaps between two CCDs, and thus only data from the EPIC-MOS1 are considered in the following analysis. 
The best determined source position with the {\sc edetect$\_$chain} tool is 
R.A.(J2000)=17$^{\rm h}$35$^{\rm m}$18\fs48 and Dec.(J2000)=-32\degr54\arcmin30\farcs2, compatible with that determined previously with \chan\ 
(see Sect.~\ref{sec:jD})\footnote{The source is relatively faint for the MOS1, and a position uncertainty  
as large as 4\arcsec can be expected in this case \citep[see discussion in][]{pavan11}.}.    
We extracted the spectrum of S2 from a circular region centered 
on the \chan\ position of the source \citep{tomsick09}. Due to the limited statistics of the data, 
the spectrum was grouped to have at least 5 counts per bin\footnote{We performed this minimal grouping to avoid issues related to empty channels, see 
heasarc.nasa.gov/xanadu/xspec/XspecManual.pdf. A check was carried out {\it a posteriori} to verify that this did not affect the results.} 
and fit with the C-statistics \citep{cash79} by using an absorbed 
power-law model. We fixed the absorption column density to the Galactic value expected in the direction of the source 
\citep[$N_H$$\simeq$1.4$\times$10$^{22}$~\textrm{cm}$^{-2}$,][]{map1, dickey90} and measured a power-law photon index of 
$\Gamma$=0.9$\pm$0.3 (C-statistics/d.o.f.=29.4/36). The corresponding 0.5--10~keV observed flux was 2.4$\times$10$^{-13}$~\ferg.
No significant evidence for coherent modulations was found in the EPIC data.

\subsection{ \jE\ }
\label{sec:jE-xmm}

\begin{figure}
\centering
\includegraphics[scale=0.36,angle=-90]{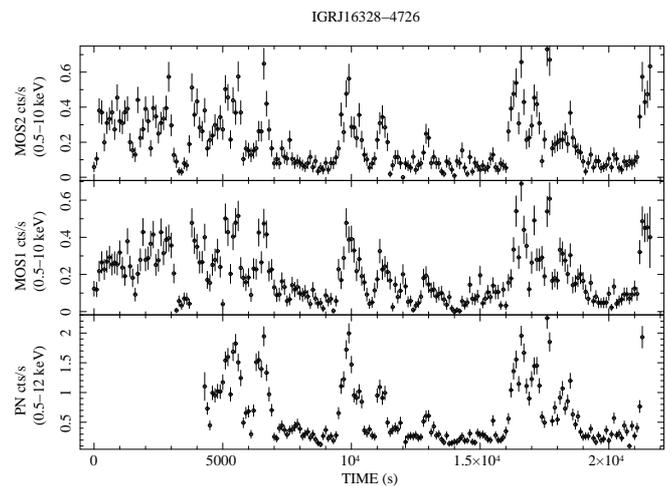}
\caption{ \xmm\ background-subtracted lightcurve of \jE.\ Data are shown in the 0.5--12~keV and 0.5--10~keV 
energy range for the EPIC-pn and EPIC-MOS, respectively. The time bin in all cases 
is 100~s.}    
\label{fig:jElcurve} 
\end{figure}

\begin{figure}
\centering
\includegraphics[scale=0.36,angle=-90]{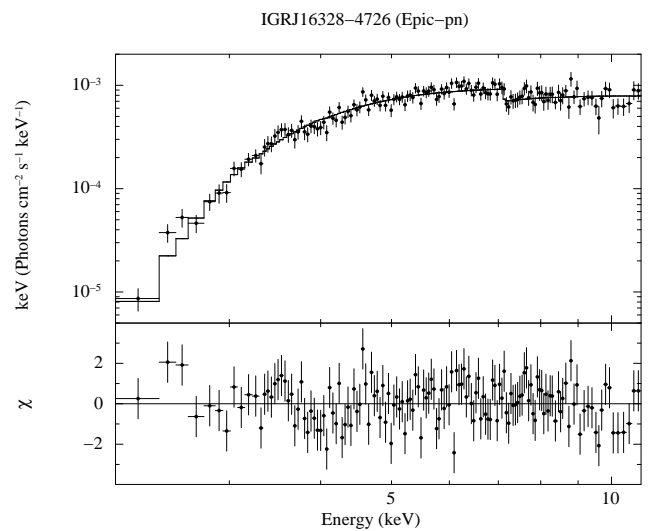}
\caption{Average \xmm\ spectrum of \jE\ extracted from the EPIC-pn data. The best-fit 
model is obtained with an absorbed power law (see text for details). The residuals from the 
fit are shown in the bottom panel.}    
\label{fig:jEspectrum} 
\end{figure}

\begin{figure}
\centering
\includegraphics[scale=0.36,angle=-90]{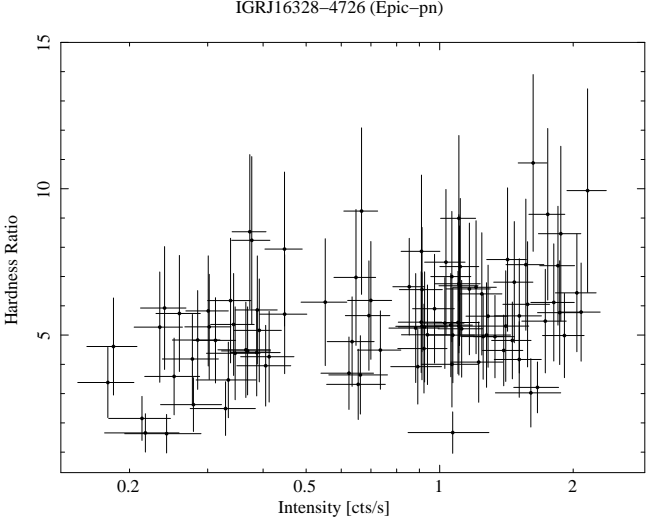}
\caption{Hardness-intensity diagram of \jE\ realized by using EPIC-pn data. For this plot, 
data were rebinned in order to obtain in each bin a S/N$\gtrsim$5.}     
\label{fig:jEHID} 
\end{figure}

\begin{figure}
\centering
\includegraphics[scale=0.35,angle=-90]{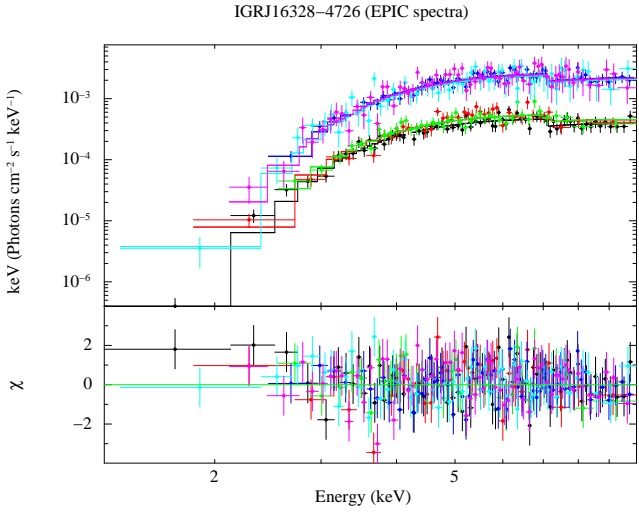}
\caption{ EPIC-pn and EPIC-MOS spectra of \jE.\ The higher three spectra were extracted by selecting only the time intervals 
in which the EPIC-pn source count-rate was $>$1.5~cts/s (0.5--12keV, see Fig.~\ref{fig:jElcurve}). The lower three spectra 
were extracted during time intervals in which the count-rate was $<$0.7~cts/s (same energy band). GTIs from the EPIC-pn 
were used also to extract the corresponding EPIC-MOS spectra. The best-fit model (solid line; $\chi^2_{\rm red}$/d.o.f.=0.9/332) is obtained with 
an absorbed power-law (the residuals from the fit are shown in the bottom panel). 
We obtained for the highest spectra N$_{\rm H}$=(17.4$\pm$1.7)$\times$10$^{22}$~cm$^{-2}$ and $\Gamma$=1.6$\pm$0.2; for the lower spectra 
we obtained N$_{\rm H}$=(18.2$\pm$1.8)$\times$10$^{22}$~cm$^{-2}$ and $\Gamma$=1.5$\pm$0.2. The corresponding observed 2--10~keV X-ray fluxes were 
2.2$\times$10$^{-11}$~erg/cm$^2$/s and 4$\times$10$^{-12}$~erg/cm$^2$/s, respectively.}   
\label{fig:jE-PNMOS} 
\end{figure}

\jE\ was observed by \xmm\ on 2011 February 20 for a total exposure time of $\sim$22~ks. 
Filtering for the flaring background intervals resulted in an effective exposure time of 
14.7~ks for the EPIC-pn and 21~ks for the two MOS cameras. 
The best determined source position with the {\sc edetect$\_$chain} tool is 
R.A.(J2000)=16$^{\rm h}$32$^{\rm m}$37\fs68 and Dec.(J2000)=-47\degr23\arcmin39.48\arcsec, compatible with that determined previously with \swift\ 
\citep{grupe09}.    
The lightcurves extracted from the three instruments and corrected for the background are  
shown in Fig.~\ref{fig:jElcurve}.  The source displayed a clear variability, with relatively small flares 
($\Delta_{L_{\rm X}}$$\lesssim$10) occurring during periods of low-level X-ray activity. The average EPIC-pn spectrum extracted 
during the observation is shown in Fig.~\ref{fig:jEspectrum}. The best fit to this spectrum ($\chi^2_{\rm red}$/d.o.f.=1.06/105) 
is obtained by using an absorbed power-law model. The measured absorption column density 
is $N_{\rm H}$=(17.5$\pm$1.0)$\times$10$^{22}$~cm$^{-2}$, and the power-law photon 
index is $\Gamma$=1.5$\pm$0.1. The average unabsorbed 2--10~keV flux 
is 1.7$\times$10$^{-11}$~erg/cm$^2$/s. We searched for possible spectral variations during the rise and decay of the small 
flares by using the hardness-intensity diagram (HID) of the source (see Fig.~\ref{fig:jEHID}). 
The Spearman rank correlation coefficient calculated for the 97 data points comprised in the plot is 0.39, 
thus indicating a statistically significant correlation between the HR and the source intensity 
(the corresponding null hypothesis probability is $\lesssim$10$^{-4}$).    
We carried out a count-rate resolved spectral analysis to check this result, but did not find any significant variation 
of the spectral parameters ($N_{\rm H}$, $\Gamma$) with the source intensity. This can be ascribed also to the 
relatively low statistics of the data. We checked that this result could not be improved by fitting simultaneously 
pn and MOS data extracted during the same time intervals (we used the good time intervals, GTIs, of the EPIC-pn to extract simultaneous spectra 
from the two MOS cameras). The results of this analysis are shown in Fig.~\ref{fig:jE-PNMOS}. We report there only the results corresponding to one 
particular selection of the source high and low count-rate time intervals. Different choices of these intervals did not lead to significant 
changes of the results.  

As shown in Fig.~\ref{fig:jElcurve}, the two EPIC-MOS cameras began observing the source about 4~ks 
earlier than the EPIC-pn. Beside the small flares visible from all the instruments, the EPIC-MOS showed 
an intriguing drop of the source count-rate $\sim$3~ks after the beginning of the observation. We performed a spectral analysis 
of the combined MOS1 and MOS2 data extracted during this event\footnote{See http://xmm.esac.esa.int/sas/current/documentation/threads/}, 
but the relatively low statistics did not permit to carry out any detailed spectral analysis (only 60 counts were recorded during the 
$\sim$0.7~ks corresponding to the event). The derived spectral parameters were fully consistent with those 
measured from the average spectrum, in turns compatible (to within the errors) with those already obtained from the 
spectral fit to the EPIC-pn data.  The highest and lowest unabsorbed fluxes of \jE\ recorded by the EPIC cameras were 
5.2$\times$10$^{-11}$~\ferg\ and 6.4$\times$10$^{-12}$~\ferg,\ respectively. 
No significant evidence for coherent modulations was found in the EPIC data of the source.  

\subsection{ \jC\ }
\label{sec:jC-xmm}

The \xmm\ observation of \jC\ reported here was performed on 2010 March 21 starting from 12:20 UTC  
for a total exposure time of 45~ks. The EPIC-pn camera was operated in Full Frame mode, while the EPIC-MOS1 
and EPIC-MOS2 were operated in Small Window and Fast Uncompressed mode, respectively. 
The source was not detected by the EPIC cameras. To estimate an upper limit on the source X-ray flux, 
we first filtered the EPIC data for the high flaring background and then applied the same technique as described 
in Sect.~\ref{sec:jD-xmm}. As the observation suffered from a very high level of flaring background, the effective exposure time 
available for the EPIC-pn was 4.0~ks. We estimated from these data a 3~$\sigma$ upper 
limit on the source X-ray count-rate of 0.0052~cts/s, and converted this value into a 0.5--10~keV unabsorbed flux of 
3.0$\times$10$^{-13}$~erg/cm$^2$/s by assuming the same spectral model used in \citet{bozzo08b}. This allows an easier comparison 
with the previous \xmm\ non-detection of the source (see Sect.~\ref{sec:jC}). We also checked that 
a less conservative selection of GTIs for the EPIC-pn data would not affect significantly the derived 
upper limit on the X-ray flux of \jC.\  

According to the ephemeris given by \citet{bird09}, the epochs of the present \xmm\ observation of \jC\ corresponds 
to phase 0.53$\pm$0.18. Even though the associated uncertainty is rather large, this result suggests that the \xmm\ observation took 
place around the system apastron. The upper limit on the flux is indeed comparable with that estimated by the previous non-detection 
of the source at phase 0.51 \citep{bozzo08b}.

\subsection{ \jB\ }
\label{sec:jB-xmm}

\jB\ was observed by \xmm\ on 2011 March 2 for a total exposure time of 
20~ksec. The three EPIC detectors were operated in Full Frame with medium filter.

\begin{figure}
\centering
\includegraphics[scale=0.35]{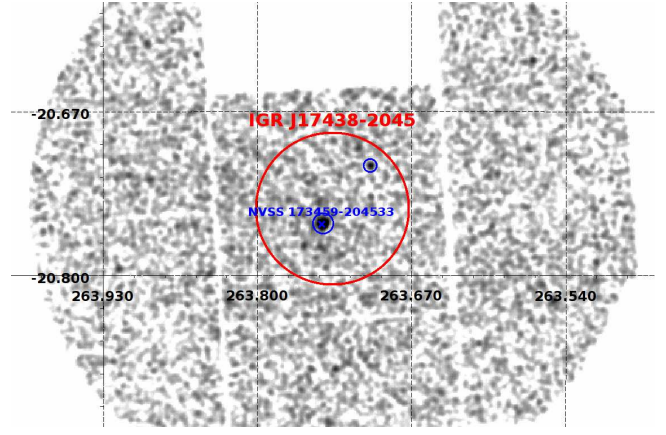}
\caption{ \xmm\ image of the FOV around \jB\ (EPIC-MOS image in the 0.5--10~keV energy band). The two soft X-ray counterparts of \jB\ 
detected within the \inte\ error circle of the source are indicated with blue circles. The radio object NVSS\,J173459$-$204533, spatially coincident 
with XMMU\,J173458.8$-$204530, is also indicated with a blue cross.}    
\label{fig:jBima} 
\end{figure}

\begin{figure}
\centering
\includegraphics[scale=0.35,angle=-90]{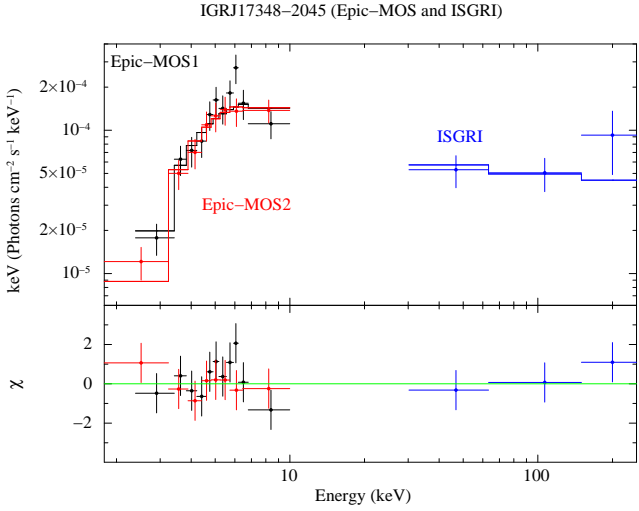}
\caption{EPIC-MOS and ISGRI spectra of XMMU\,J173458.8$-$204530, the most likely counterpart to \jB.\ The best fit is obtained 
by using an absorbed power-law model (see Sect.~\ref{sec:jB-xmm} for details).}    
\label{fig:jBspectra} 
\end{figure}

The observation was affected by a high flaring background, resulting in an effective exposure time of 10~ks for the two 
MOS cameras and 4~ks for pn. We searched simultaneously for the detected sources in all the cleaned EPIC images by using the \verb+edetect_chain+ {\em SAS} tool.  
Inside the \inte\ error circle around \jB\ we found two soft X-ray sources located at 
\mbox{$RA=17^{\mathrm{h}}34^{\mathrm{m}}58.80^{\mathrm{s}}$}, 
\mbox{$Dec=-20$\degr45\arcmin30.96\arcsec}, and 
\mbox{$RA=17^{\mathrm{h}}34^{\mathrm{m}}49.20^{\mathrm{s}}$}, 
\mbox{$Dec=-20$\degr42\arcmin44.64\arcsec}, 
respectively. 
According to the \xmm\ convention, we named the two sources XMMU\,J173458.8$-$204530 and XMMU\,J173449.2$-$204244. 
The former source is clearly visible in all the images of the individual instruments and located close to the center of the \inte\ error-circle; the latter 
is detected only when all EPIC images are used simultaneously and located at the very rim of the \inte\ error circle (see Fig.~\ref{fig:jBima}). 
For this latter source no meaningful spectral information could be extracted, and we thus assume in the following 
that XMMU\,J173458.8$-$204530 is the true counterpart to \jB.\ 
For this source, the statistic was enough to perform a spectral analysis. We extracted the three 
MOS and pn spectra using a circular extraction region centered on the best determined source position (see above). 
The background spectrum was extracted from the closest source-free region to XMMU\,J173458.8$-$204530 
(we checked that different choices of the background extraction region did not affect significantly the spectral results).
On the EPIC-pn the source was located on the gap between two CCDs, and thus no spectrum could be extracted. 
A simultaneous fit to the EPIC-MOS spectra with an absorbed power-law model gave 
$N_{\rm H}$=17$^{+7}_{-5}$$\times$10$^{22}$~cm$^{-2}$, $\Gamma$=1.5$^{+0.8}_{-0.7}$ ($\chi^2_{\rm red}$/d.o.f.= 0.7/16). 
The estimated absorption column density is much higher than the Galactic value expected in the direction of the source 
\citep[$\sim$2$\times$10$^{21}$~cm$^{-2}$;][]{dickey90}.  
The unabsorbed 2--10~keV flux is $F_{\rm 2-10~keV}$=2.2$\times$10$^{-12}$\ferg.\  No significant evidence for coherent 
modulations was found in the EPIC data of \jB.\  
For this source we also retrieved the long-term ISGRI spectrum from the {\sc Heavens} on-line tool (effective exposure time 3.9~Ms) 
and performed a combined fit with the MOS spectra (see Fig.~\ref{fig:jBspectra}). A normalization constant with respect to MOS was included 
in the fit to account for the inter-calibration between the instruments and possible variability of the source. The best fit model 
($\chi^2_{\rm red}$/d.o.f.= 0.7/18) was obtained using the same absorbed power-law model as described above, with values of the absorption column density 
and power-law photon index fully compatible to within the uncertainties (no significant improvement on the uncertainties of the spectral parameters could be obtained). 
The normalization constant is $0.4_{-0.3}^{+0.9}$, thus supporting the idea that 
the source is a persistent hard X-ray emitter (we verified that fixing the normalization constant to 1 for both the ISGRI and MOS spectra would not 
affect significantly the results of the fit). 
The estimated 20--40~keV and 40--100~keV X-ray fluxes were 2$\times$10$^{-12}$~\ferg\ and 5.2$\times$10$^{-12}$~\ferg,\ respectively. These are compatible 
with the values reported by \citet{bird09}. 
We also searched for counterparts to XMMU\,J173458.8$-$204530 in the optical, infrared and radio domain. 
The closest catalogued objects to the XMM source in the optical and infrared domains are USNO-B1.0\,0692-0466331 (B1=15.82~mag, R1=12.97~mag, 
B2=15.40~mag, R2=14.35~mag, I=11.70~mag) and 2MASS\,J17345863$-$2045292 (J=10.532~mag, H=9.446~mag, K=9.019~mag), respectively. 
These two objects are located at the rim of the \xmm\ error circle. In addition, the large absorption measured 
in the X-ray domain, which corresponds to an optical extinction of $\simeq$8~mag according to \citet{guver09}, suggests that an optical counterpart 
might not be expected. Further observations are needed to confirm these associations (e.g. with \chan\ ).
From the 1.4~GHz continuum NRAO Very Large Sky Survey \citep[NVSS][]{nvss}, we also noticed that the object 
NVSS\,J173459$-$204533 lies relatively close to the XMM source ($\sim$4.8''). 
The best determined position of the radio object is at \mbox{$RA=17^{\mathrm{h}}34^{\mathrm{m}}59.08^{\mathrm{s}}$} ($\pm$0.09~s), 
\mbox{$Dec=-20$\degr45\arcmin33.8\arcsec} ($\pm$1.2~arcsec), and the integrated flux density at 1.4~GHz is $13.2 \pm 0.6$~mJy \citep{nvss}.

\section{Discussion and conclusions}
\label{sec:discussion}

We reported here on the observations  
of five hard X-ray emitters obtained with \xmm.\ 
Among these, four (\jA,\ \jD,\ \jE,\ \jC) are  
HMXBs, whereas \jB\ is still unidentified.

\subsection{ \jA\ } 
\label{sec:jAdiscussion}

The \xmm\ observation of this source revealed a timing and spectral behaviour 
that are typical of classical SGXBs. The EPIC-pn lightcurve shows a relatively moderate variability on time 
scales of hundreds seconds with variations in the X-ray flux of a factor $\sim$3-4. 
These variations are usually ascribed to changes in the mass accretion rate 
due to instabilities and density fluctuations in the supergiant wind (see Sect.~\ref{sec:intro}). 
The EPIC spectra could not be fit with a simple absorbed power-law model, and clearly showed the 
presence of an excess in the residuals below 2--3~keV. The presence of this ``soft excess'', requiring 
an additional spectral component to obtain an acceptable fit to the data, is believed to be 
a ubiquitous feature of binary systems hosting accreting NSs \citep{hickox04}. Its detectability is mostly related to 
the absorption column density in the direction of the source and its flux. 
In wind accreting binaries with X-ray luminosities comparable to that of \jA\ 
($\simeq$3$\times$10$^{34}$~erg/s, assuming a distance of 6.1~kpc), the soft component 
is most likely originated by thermal X-ray photons close to the NS surface 
or by photoionized or collisionally heated diffuse gas in the binary system \citep{hickox04}. 
 
The statistics of the EPIC spectra below 2--3~keV shown in Sect.~\ref{sec:jA-xmm} were unfortunately too low to clearly distinguish 
between these possibilities. An acceptable fit could be obtained by using either  
a hot ($kT$$\simeq$1.2~keV) compact ($\simeq$0.1~km) BB component, or a colder ($kT$$\simeq$0.1~keV) and more extended 
($\sim$130~km) thermal emission. The latter model would probably be more plausible in the case of \jA\ as low luminosity 
wind accreting systems are not expected to display hot emitting spots on the NS surface \citep[see discussion in][and references therein]{bozzo10}. 
As an alternative interpretation, we also showed in Table~\ref{tab:jAspec} that a partial-covering model would provide an acceptable 
description of the X-ray spectrum. In this case, the soft excess would be produced by the effect 
of partial obscuration of the emission from the NS by the surrounding high density material 
\citep[see e.g.,][]{tomsick09b}.  

The combined fit of the EPIC spectra with that obtained from the long-term monitoring of the source with ISGRI, revealed 
that the models described above can also account for the higher energy emission recorded from the source (up to $\sim$60~keV). 
The normalization constants between all the instruments were compatible with unity, thus suggesting that 
\jA\ is characterized by a virtually constant persistent X-ray flux. This is expected for a classical SGXB.

\subsection{ \jD\ }

The \xmm\ observation detected only one of the two possible counterparts to the \inte\ source 
identified before with \swift\ and \chan\ (see Sect.~\ref{sec:jD-xmm}). For the most likely  
counterpart (S1), the non-detection by \xmm\ sets a 3$\sigma$ upper limit of  7$\times$10$^{-14}$~\ferg\ 
on its observed 0.5--10~keV X-ray flux. This is a factor of $\sim$6 lower than the previously determined upper limit 
(see Sect.~\ref{sec:jD-xmm}). This observation thus increases significantly 
the dynamic range in the X-ray luminosity of \jD,\ supporting the classification of 
this source as a SFXT.  

We also note that the X-ray flux of the source S2 measured by \xmm\ is roughly compatible with that 
reported previously by \citet{vercellone09}, thus supporting the idea that this is a persistent object 
unrelated to the \inte\ source.

\subsection{ \jE\ } 

The \xmm\ observation of this source revealed a variability on time scales of hundreds of seconds 
(see Fig.~\ref{fig:jElcurve}) that is very reminiscent of that observed from the SFXT prototypes IGR\,J17391$-$3021 and 
IGR\,J08408$-$4503 \citep{bozzo10}. This provides further support to the association of \jE\ with this class of objects. 

At odds with some other SFXTs observed in quiescence, we could not detect any soft excess in the X-ray spectrum of the source. 
According to the discussion in Sect.~\ref{sec:jAdiscussion}, this spectral component might have gone undetected in \jE\ 
due to its relatively high absorption column density ($N_{\rm H}$$\sim$1.8$\times$10$^{23}$~cm$^{-2}$) 
and the lower exposure time with respect to the \xmm\ observations of other quiescent SFXT sources with 
a similar X-ray luminosity \citep{bozzo10}. 

As a final remark we note that the lowest X-ray flux measured by \xmm\ is compatible with the upper limit reported previously 
with \swift\,/XRT by \citet{fiocchi10}, thus suggesting that this is the real quiescent emission level of the source.

\subsection{ \jC\ } 

This source is one of the confirmed SFXTs, and has been observed in outburst several times with \inte\ and \swift.\ 
The properties of its soft X-ray emission ($\lesssim$10~keV) during outbursts were investigated in detail thanks to the 
observations performed with the XRT on-board \swift,\ but so far no observations with a focusing X-ray telescope endowed with a high 
sensitivity was able to detect the source in quiescence. The \xmm\ observation reported here and the one performed previously in 
2006 \citep{bozzo08b} were carried out close to the system apastron and provided a similar 3$\sigma$ upper limit on the source X-ray 
luminosity of $\simeq$2$\times$10$^{32}$~erg/s (assuming a distance of 2.5~kpc).

The nature of the prolonged quiescent state of \jC\ around the apastron is presently unknown, due to the lack of proper 
spectral information. 
As the orbital period of the system is $\sim$30~days, the presence of X-ray eclipses during apastron would imply relatively strong 
constraints on the system inclination. A high eccentricity ($e$) might help to reduce 
the accretion rate at the apastron, but the value of $e$ inferred for \jC\ (0.3-0.4) can hardly explain the large dynamic 
range in the X-ray luminosity displayed by the source \citep[][]{bird09,zurita09}. A further reduction of the mass accretion rate would thus require 
some additional mechanism to be at work at this orbital phase. Proposed models involve the inhibition of the accretion by the NS 
centrifugal and/or the magnetic barrier \citep{grebenev07,bozzo08b}, or the presence of a highly structured wind from the companion star which is extremely rarefied 
around apastron and permits at this phase only accretion at a very low-level \citep[see, e.g.][and references therein]{zurita09}.  
Further observations of \jC\ around orbital phase 0.4-0.6 with high sensitivity X-ray telescopes, 
like \xmm\ and \chan,\ are required to clarify the origin of the lowest 
X-ray luminosity state reached by this source.

\subsection{ \jB\ } 

This source was never observed before in the soft X-ray domain. The \xmm\ data reported here allowed us to identify the counterpart to 
the \inte\ source and revealed that the source X-ray emission is intrinsically highly absorbed ($\simeq$10$^{23}$~cm$^{-2}$, see Sect.~\ref{sec:jB-xmm}) 
and well described by a power-law model with $\Gamma$$\sim$1.5. A simultaneous fit with the long-term ISGRI spectrum also supports the idea 
that the source is a persistent hard X-ray emitter. 

The uncertainty in the EPIC-MOS position does not allow to firmly identify the counterparts at other wavelengths, 
but we remark here the possible association with the radio object  NVSS\,J173459$-$204533. If this association will be confirmed by future 
observations performed with higher spatial-resolution instruments, we propose that \jB\ could be a highly absorbed active 
Galactic nuclei (AGN). A number of these sources were recently discovered with \inte\ \citep[see, e.g.][and references therein]{ricci11}. 
Further multi-wavelength observations of this source are required in order to 
establish its real nature.

\section*{Acknowledgments}
EB thanks A. Tramacere and C. Ricci for useful discussions, and an anonymous referee 
for useful comments. 
The results presented in this paper are based on observations obtained 
with \xmm,\ an ESA science mission with instruments and contributions 
directly funded by ESA Member States and NASA. This work also
makes use of observations taken by \inte,\ an ESA project
with instruments and science data centre funded by ESA member
states (especially the PI countries: Denmark, France, Germany, Italy,
Switzerland, Spain), Poland and with the participation of Russia and
the USA.

\bibliographystyle{aa}
\bibliography{HMXBs}

\begin{thebibliography}{46}
\expandafter\ifx\csname natexlab\endcsname\relax\def\natexlab#1{#1}\fi

\bibitem[{{Bird} {et~al.}(2010){Bird}, {Bazzano}, {Bassani}, {Capitanio},
  {Fiocchi}, {Hill}, {Malizia}, {McBride}, {Scaringi}, {Sguera}, {Stephen},
  {Ubertini}, {Dean}, {Lebrun}, {Terrier}, {Renaud}, {Mattana}, {G{\"o}tz},
  {Rodriguez}, {Belanger}, {Walter}, \& {Winkler}}]{bird10}
{Bird}, A.~J., {Bazzano}, A., {Bassani}, L., {et~al.} 2010, \apjs, 186, 1

\bibitem[{{Bird} {et~al.}(2009){Bird}, {Bazzano}, {Hill}, {McBride}, {Sguera},
  {Shaw}, \& {Watkins}}]{bird09}
{Bird}, A.~J., {Bazzano}, A., {Hill}, A.~B., {et~al.} 2009, \mnras, 393, L11

\bibitem[{{Bird} {et~al.}(2007){Bird}, {Malizia}, {Bazzano}, {Barlow},
  {Bassani}, {Hill}, {B{\'e}langer}, {Capitanio}, {Clark}, {Dean}, {Fiocchi},
  {G{\"o}tz}, {Lebrun}, {Molina}, {Produit}, {Renaud}, {Sguera}, {Stephen},
  {Terrier}, {Ubertini}, {Walter}, {Winkler}, \& {Zurita}}]{bird07}
{Bird}, A.~J., {Malizia}, A., {Bazzano}, A., {et~al.} 2007, \apjs, 170, 175

\bibitem[{{Bozzo} {et~al.}(2008{\natexlab{a}}){Bozzo}, {Campana}, {Stella},
  {Falanga}, {Israel}, {Rampy}, {Smith}, \& {Negueruela}}]{bozzo08c}
{Bozzo}, E., {Campana}, S., {Stella}, L., {et~al.} 2008{\natexlab{a}}, The
  Astronomer's Telegram, 1493

\bibitem[{{Bozzo} {et~al.}(2008{\natexlab{b}}){Bozzo}, {Falanga}, \&
  {Stella}}]{bozzo08b}
{Bozzo}, E., {Falanga}, M., \& {Stella}, L. 2008{\natexlab{b}}, \apj, 683, 1031

\bibitem[{{Bozzo} {et~al.}(2011){Bozzo}, {Giunta}, {Cusumano}, {Ferrigno},
  {Walter}, {Campana}, {Falanga}, {Israel}, \& {Stella}}]{bozzo11}
{Bozzo}, E., {Giunta}, A., {Cusumano}, G., {et~al.} 2011, \aap, 531, A130

\bibitem[{{Bozzo} {et~al.}(2010){Bozzo}, {Stella}, {Ferrigno}, {Giunta},
  {Falanga}, {Campana}, {Israel}, \& {Leyder}}]{bozzo10}
{Bozzo}, E., {Stella}, L., {Ferrigno}, C., {et~al.} 2010, \aap, 519, A6

\bibitem[{{Bozzo} {et~al.}(2008{\natexlab{c}}){Bozzo}, {Stella}, {Israel},
  {Falanga}, \& {Campana}}]{bozzo08}
{Bozzo}, E., {Stella}, L., {Israel}, G., {Falanga}, M., \& {Campana}, S.
  2008{\natexlab{c}}, \mnras, 391, L108

\bibitem[{{Bulgarelli} {et~al.}(2009){Bulgarelli}, {Gianotti}, {Trifoglio},
  {Striani}, {Tavani}, {Sabatini}, {Vercellone}, {Feroci}, {Lazzarotto}, {Del
  Monte}, {Pittori}, {Verrecchia}, {Pellizzoni}, {Pilia}, {Chen}, {Giuliani},
  {D'Ammando}, {Piano}, {Pucella}, {Vittorini}, {Costa}, {Donnarumma},
  {Pacciani}, {Soffitta}, {Evangelista}, {Lapshov}, {Rapisarda}, {Argan},
  {Trois}, {de Paris}, {Marisaldi}, {Di Cocco}, {Labanti}, {Fuschino}, {Galli},
  {Caraveo}, {Mereghetti}, {Perotti}, {Fiorini}, {Zambra}, {Barbiellini},
  {Longo}, {Moretti}, {Vallazza}, {Picozza}, {Morselli}, {Prest}, {Lipari},
  {Zanello}, {Cattaneo}, {Santolamazza}, {Colafrancesco}, {Giommi}, \&
  {Salotti}}]{bulgarelli09}
{Bulgarelli}, A., {Gianotti}, F., {Trifoglio}, M., {et~al.} 2009, The
  Astronomer's Telegram, 2017, 1

\bibitem[{{Cash}(1979)}]{cash79}
{Cash}, W. 1979, \apj, 228, 939

\bibitem[{{Condon} {et~al.}(1998){Condon}, {Cotton}, {Greisen}, {Yin},
  {Perley}, {Taylor}, \& {Broderick}}]{nvss}
{Condon}, J.~J., {Cotton}, W.~D., {Greisen}, E.~W., {et~al.} 1998, \aj, 115,
  1693

\bibitem[{{Corbet} {et~al.}(2010){Corbet}, {Barthelmy}, {Baumgartner}, {Krimm},
  {Markwardt}, {Skinner}, \& {Tueller}}]{corbet10}
{Corbet}, R.~H.~D., {Barthelmy}, S.~D., {Baumgartner}, W.~H., {et~al.} 2010,
  The Astronomer's Telegram, 2588

\bibitem[{{Cusumano} {et~al.}(2010){Cusumano}, {La Parola}, {Segreto},
  {Ferrigno}, {Maselli}, {Sbarufatti}, {Romano}, {Chincarini}, {Giommi},
  {Masetti}, {Moretti}, {Parisi}, \& {Tagliaferri}}]{cusumano10}
{Cusumano}, G., {La Parola}, V., {Segreto}, A., {et~al.} 2010, \aap, 524, A64

\bibitem[{{D'A{\`i}} {et~al.}(2011){D'A{\`i}}, {La Parola}, {Cusumano},
  {Segreto}, {Romano}, {Vercellone}, \& {Robba}}]{dai11}
{D'A{\`i}}, A., {La Parola}, V., {Cusumano}, G., {et~al.} 2011, \aap, 529, A30

\bibitem[{{Dickey} \& {Lockman}(1990)}]{dickey90}
{Dickey}, J.~M. \& {Lockman}, F.~J. 1990, \araa, 28, 215

\bibitem[{{Fiocchi} {et~al.}(2010){Fiocchi}, {Sguera}, {Bazzano}, {Bassani},
  {Bird}, {Natalucci}, \& {Ubertini}}]{fiocchi10}
{Fiocchi}, M., {Sguera}, V., {Bazzano}, A., {et~al.} 2010, \apjl, 725, L68

\bibitem[{{Grebenev} \& {Sunyaev}(2007)}]{grebenev07}
{Grebenev}, S.~A. \& {Sunyaev}, R.~A. 2007, Astronomy Letters, 33, 149

\bibitem[{{Grupe} {et~al.}(2009){Grupe}, {Kennea}, {Evans}, {Romano},
  {Markwardt}, \& {Chester}}]{grupe09}
{Grupe}, D., {Kennea}, J., {Evans}, P., {et~al.} 2009, The Astronomer's
  Telegram, 2075

\bibitem[{{G{\"u}ver} \& {{\"O}zel}(2009)}]{guver09}
{G{\"u}ver}, T. \& {{\"O}zel}, F. 2009, \mnras, 400, 2050

\bibitem[{{Hickox} {et~al.}(2004){Hickox}, {Narayan}, \& {Kallman}}]{hickox04}
{Hickox}, R.~C., {Narayan}, R., \& {Kallman}, T.~R. 2004, \apj, 614, 881

\bibitem[{{in 't Zand} {et~al.}(1998){in 't Zand}, {Heise}, {Smith}, {Muller},
  {Ubertini}, \& {Bazzano}}]{zand98}
{in 't Zand}, J., {Heise}, J., {Smith}, M., {et~al.} 1998, \iaucirc, 6840, 2

\bibitem[{{in 't Zand} {et~al.}(2006){in 't Zand}, {Jonker}, {Mendez}, \&
  {Markwardt}}]{zand06}
{in 't Zand}, J., {Jonker}, P., {Mendez}, M., \& {Markwardt}, C. 2006, The
  Astronomer's Telegram, 915

\bibitem[{{in 't Zand}(2005)}]{zand05}
{in 't Zand}, J.~J.~M. 2005, \aap, 441, L1

\bibitem[{{Kalberla} {et~al.}(2005){Kalberla}, {Burton}, {Hartmann}, {Arnal},
  {Bajaja}, {Morras}, \& {P{\"o}ppel}}]{map1}
{Kalberla}, P.~M.~W., {Burton}, W.~B., {Hartmann}, D., {et~al.} 2005, \aap,
  440, 775

\bibitem[{{Kreykenbohm} {et~al.}(2008){Kreykenbohm}, {Wilms}, {Kretschmar},
  {Torrej{\'o}n}, {Pottschmidt}, {Hanke}, {Santangelo}, {Ferrigno}, \&
  {Staubert}}]{kreykenbohm11}
{Kreykenbohm}, I., {Wilms}, J., {Kretschmar}, P., {et~al.} 2008, \aap, 492, 511

\bibitem[{{Kuulkers} {et~al.}(2006){Kuulkers}, {Shaw}, {Paizis}, {Gros},
  {Chenevez}, {Sanchez-Fernandez}, {Brandt}, {Courvoisier}, {Garau}, {Ebisawa},
  {Kretschmar}, {Markwardt}, {Mowlavi}, {Oosterbroek}, {Orr}, {Oneca}, \&
  {Wijnands}}]{kuulkers06}
{Kuulkers}, E., {Shaw}, S., {Paizis}, A., {et~al.} 2006, The Astronomer's
  Telegram, 874

\bibitem[{{Malizia} {et~al.}(2011){Malizia}, {Landi}, {Bassani}, {Bird},
  {Gehrels}, \& {Kennea}}]{maliziaatel}
{Malizia}, A., {Landi}, R., {Bassani}, L., {et~al.} 2011, The Astronomer's
  Telegram, 3294

\bibitem[{{Masetti} {et~al.}(2010){Masetti}, {Parisi}, {Palazzi},
  {Jim{\'e}nez-Bail{\'o}n}, {Chavushyan}, {Bassani}, {Bazzano}, {Bird}, {Dean},
  {Charles}, {Galaz}, {Landi}, {Malizia}, {Mason}, {McBride}, {Minniti},
  {Morelli}, {Schiavone}, {Stephen}, \& {Ubertini}}]{masetti10}
{Masetti}, N., {Parisi}, P., {Palazzi}, E., {et~al.} 2010, \aap, 519, A96

\bibitem[{{Negueruela}(2010)}]{negueruela10}
{Negueruela}, I. 2010, in Astronomical Society of the Pacific Conference
  Series, Vol. 422, High Energy Phenomena in Massive Stars, ed.
  {J.~Mart{\'{\i}}, P.~L.~Luque-Escamilla, \& J.~A.~Combi}, 57

\bibitem[{{Negueruela} {et~al.}(2006){Negueruela}, {Smith}, {Reig}, {Chaty}, \&
  {Torrej{\'o}n}}]{negueruela06}
{Negueruela}, I., {Smith}, D.~M., {Reig}, P., {Chaty}, S., \& {Torrej{\'o}n},
  J.~M. 2006, in ESA Special Publication, Vol. 604, The X-ray Universe 2005,
  ed. A.~{Wilson}, 165

\bibitem[{{Pavan} {et~al.}(2011){Pavan}, {Bozzo}, {Ferrigno}, {Ricci},
  {Manousakis}, {Walter}, \& {Stella}}]{pavan11}
{Pavan}, L., {Bozzo}, E., {Ferrigno}, C., {et~al.} 2011, \aap, 526, A122

\bibitem[{{Reig}(2011)}]{reig11}
{Reig}, P. 2011, \apss, 332, 1

\bibitem[{{Ricci} {et~al.}(2011){Ricci}, {Walter}, {Courvoisier}, \&
  {Paltani}}]{ricci11}
{Ricci}, C., {Walter}, R., {Courvoisier}, T.~J.-L., \& {Paltani}, S. 2011,
  \aap, 532, A102

\bibitem[{{Romano} {et~al.}(2011){Romano}, {La Parola}, {Vercellone},
  {Cusumano}, {Sidoli}, {Krimm}, {Pagani}, {Esposito}, {Hoversten}, {Kennea},
  {Page}, {Burrows}, \& {Gehrels}}]{romano11}
{Romano}, P., {La Parola}, V., {Vercellone}, S., {et~al.} 2011, \mnras, 410,
  1825

\bibitem[{{Sguera} {et~al.}(2006){Sguera}, {Bazzano}, {Bird}, {Dean},
  {Ubertini}, {Barlow}, {Bassani}, {Clark}, {Hill}, {Malizia}, {Molina}, \&
  {Stephen}}]{sguera06}
{Sguera}, V., {Bazzano}, A., {Bird}, A.~J., {et~al.} 2006, \apj, 646, 452

\bibitem[{{Sguera} {et~al.}(2011){Sguera}, {Drave}, {Bird}, {Bazzano}, {Landi},
  \& {Ubertini}}]{sguera11}
{Sguera}, V., {Drave}, S.~P., {Bird}, A.~J., {et~al.} 2011, \mnras, 417, 573

\bibitem[{{Sidoli} {et~al.}(2009){Sidoli}, {Romano}, {Esposito}, {Parola},
  {Kennea}, {Krimm}, {Chester}, {Bazzano}, {Burrows}, \& {Gehrels}}]{sidoli09}
{Sidoli}, L., {Romano}, P., {Esposito}, P., {et~al.} 2009, \mnras, 400, 258

\bibitem[{{Smith} {et~al.}(1998){Smith}, {Main}, {Marshall}, {Swank}, {Heindl},
  {Leventhal}, {in 't Zand}, \& {Heise}}]{smith98}
{Smith}, D.~M., {Main}, D., {Marshall}, F., {et~al.} 1998, \apjl, 501, L181

\bibitem[{{Stella} {et~al.}(1986){Stella}, {White}, \& {Rosner}}]{stella86}
{Stella}, L., {White}, N.~E., \& {Rosner}, R. 1986, \apj, 308, 669

\bibitem[{{Tomsick} {et~al.}(2009{\natexlab{a}}){Tomsick}, {Chaty},
  {Rodriguez}, {Walter}, \& {Kaaret}}]{tomsick09}
{Tomsick}, J.~A., {Chaty}, S., {Rodriguez}, J., {Walter}, R., \& {Kaaret}, P.
  2009{\natexlab{a}}, \apj, 701, 811

\bibitem[{{Tomsick} {et~al.}(2009{\natexlab{b}}){Tomsick}, {Chaty},
  {Rodriguez}, {Walter}, {Kaaret}, \& {Tovmassian}}]{tomsick09b}
{Tomsick}, J.~A., {Chaty}, S., {Rodriguez}, J., {et~al.} 2009{\natexlab{b}},
  \apj, 694, 344

\bibitem[{{Torrej{\'o}n} {et~al.}(2010){Torrej{\'o}n}, {Negueruela}, {Smith},
  \& {Harrison}}]{torrejon10}
{Torrej{\'o}n}, J.~M., {Negueruela}, I., {Smith}, D.~M., \& {Harrison}, T.~E.
  2010, \aap, 510, A61

\bibitem[{{Vercellone} {et~al.}(2009){Vercellone}, {D'Ammando}, {Striani},
  {Tavani}, {Sabatini}, {Bulgarelli}, {Gianotti}, {Trifoglio}, {Feroci},
  {Lazzarotto}, {Del Monte}, {Pittori}, {Verrecchia}, {Pellizzoni}, {Pilia},
  {Chen}, {Giuliani}, {Piano}, {Pucella}, {Vittorini}, {Costa}, {Donnarumma},
  {Pacciani}, {Soffitta}, {Evangelista}, {Lapshov}, {Rapisarda}, {Argan},
  {Trois}, {de Paris}, {Marisaldi}, {Di Cocco}, {Labanti}, {Fuschino}, {Galli},
  {Caraveo}, {Mereghetti}, {Perotti}, {Fiorini}, {Zambra}, {Barbiellini},
  {Longo}, {Moretti}, {Vallazza}, {Picozza}, {Morselli}, {Prest}, {Lipari},
  {Zanello}, {Cattaneo}, {Rappoldi}, {Santolamazza}, {Colafrancesco}, {Giommi},
  {Salotti}, {Romano}, {Burrows}, \& {Gehrels}}]{vercellone09}
{Vercellone}, S., {D'Ammando}, F., {Striani}, E., {et~al.} 2009, The
  Astronomer's Telegram, 2019

\bibitem[{{Walter} \& {Zurita Heras}(2007)}]{walter07}
{Walter}, R. \& {Zurita Heras}, J. 2007, \aap, 476, 335

\bibitem[{{Yamauchi} {et~al.}(1995){Yamauchi}, {Aoki}, {Hayashida}, {Kaneda},
  {Koyama}, {Sugizaki}, {Tanaka}, {Tomida}, \& {Tsuboi}}]{yamauchi95}
{Yamauchi}, S., {Aoki}, T., {Hayashida}, K., {et~al.} 1995, \pasj, 47, 189

\bibitem[{{Zurita Heras} \& {Chaty}(2009)}]{zurita09}
{Zurita Heras}, J.~A. \& {Chaty}, S. 2009, \aap, 493, L1

\end{thebibliography}
\end{document}